\documentclass[aps, prb, preprint, bibnotes, preprintnumbers, amsmath, amssymb, superscriptaddress]{revtex4-1}
\usepackage{float}

\usepackage{graphicx}
\usepackage{dcolumn}
\usepackage{bm}

\usepackage{color}

\begin{document}

\title{Enhancement of spin mixing conductance in La$_{0.7}$Sr$_{0.3}$MnO$_{3}$/LaNiO$_{3}$/SrRuO$_{3}$ heterostructures}

\author{Christoph Hauser}
\affiliation{Institute of Physics,
Martin-Luther University Halle-Wittenberg, Von-Danckelmann-Platz 3,
06120 Halle, Germany}

\author{Camillo Ballani}
\affiliation{Institute of Physics, Martin-Luther University
Halle-Wittenberg, Von-Danckelmann-Platz 3, 06120 Halle, Germany}

\author{Philipp D\"urrenfeld}
\affiliation{Institute of Physics, Martin-Luther University
Halle-Wittenberg, Von-Danckelmann-Platz 3, 06120 Halle, Germany}

\author{Frank Heyroth}
\affiliation{Interdisziplin\"ares Zentrum f\"ur Materialwissenschaften, Martin-Luther University Halle-Wittenberg, Nanotechnikum Weinberg, Heinrich-Damerow-Str.~4, 06120 Halle, Germany}

\author{Philip Trempler}
\affiliation{Institute of Physics, Martin-Luther University
Halle-Wittenberg, Von-Danckelmann-Platz 3, 06120 Halle, Germany}

\author{Stefan G. Ebbinghaus}
\affiliation{Institute of Chemistry, Martin-Luther University
Halle-Wittenberg, Kurt-Mothes-Str. 2, 06120 Halle, Germany}

\author{Evangelos Th. Papaioannou}
\affiliation{Institute of Physics, Martin-Luther University
Halle-Wittenberg, Von-Danckelmann-Platz 3, 06120 Halle, Germany}

\author{Georg Schmidt}
\email[Correspondence to G. Schmidt:\,\,\,]{Georg.Schmidt@physik.uni-halle.de}
\affiliation{Institute of Physics,
Martin-Luther University Halle-Wittenberg, Von-Danckelmann-Platz 3,
06120 Halle, Germany}
\affiliation{Interdisziplin\"ares Zentrum f\"ur Materialwissenschaften, Martin-Luther University Halle-Wittenberg, Nanotechnikum Weinberg, Heinrich-Damerow-Str.~4, 06120 Halle, Germany}

\begin{abstract}
We investigate spin pumping and the effective spin mixing conductance in heterostructures based on magnetic oxide trilayers composed of La$_{0.7}$Sr$_{0.3}$MnO$_3$ (LSMO), LaNiO$_3$ (LNO), and SrRuO$_3$ (SRO). The heterostructures serve as a model system for an estimation of the effective spin mixing conductance at the different interfaces. Our results show that by introducing a LNO interlayer between LSMO and SRO, the total effective spin mixing conductance increases due to the much more favourable interface of LSMO/LNO with respect to the LSMO/SRO interface. Neverheless, the spin current into the SRO does not decrease because of the spin diffusion length of $\lambda_\text{LNO}\approx$\,3.3\,nm in the LNO. This value is two times higher than that of SRO. Our results show the potential of using oxide interfaces to tune the effective spin mixing conductance in heterostructures and to bring novel functionalities into spintronics by implementing complex oxides.
\end{abstract}

\flushbottom
\maketitle
\thispagestyle{empty}

\section{INTRODUCTION}
The current research on charge-to-spin current conversion effects such as the spin-Hall effect (SHE) offers great potential for applications in the field of spintronics and spin-orbitronics\cite{Sinova, Hoffmann}. Among many studies on magnetic metallic and dielectric materials for the most efficient charge-to-current conversion, oxides have attracted less attention. However, incorporating oxides into the spin current research field can be advantageous due to their tremendous variety of properties (e.g. electronic transport, magnetism) that can be tuned by deposition parameters (e.g. stoichiometry, O$_2$ pressure, strain) and that depend on the operation conditions (e.g. temperature, magnetic and electric fields). Furthermore, many oxide materials have commensurate lattice constants with perovskite-like structure that allow for very smooth interfaces in oxide heterostructures and hence can lead to well-defined properties at these interfaces.\\
In this paper, we investigate spin pumping and we calculate the effective spin mixing conductance at low temperatures in LSMO/LNO/SRO heterostructures. LSMO is a prominent oxide material with a rich phase diagram\,\cite{Cui}. We use  La$_{1-x}$Sr$_{x}$MnO$_3$ (LSMO)  with $x$\,=\,0.3 where a ferromagnetic metallic phase up to 370\,K in bulk material\,\cite{Cui} is observed. Also, the inherently bad-metallic\,\cite{Klein} oxide SRO, shows a paramagnetic to ferromagnetic transition around 155\,K\,\cite{Koster}. LNO, contrary to all other rare earth (R) nickelates RNiO$_3$ with a metal-insulator transition\,\cite{Guo}, is known to remain in a paramagnetic conducting phase even at low temperature. Previous studies on the spin pumping and the inverse spin-Hall effect in La$_{0.7}$Sr$_{0.3}$MnO$_3$/SrRuO$_{3}$ (LSMO/SRO) bilayers\,\cite{Wahler} have shown that SRO layers were acting as a spin sink exhibiting an ISHE which is similar in magnitude to that of Pt but of opposite sign. Recently, Ghosh \textit{et al.}\,\cite{Ghosh} proved the Kondo effect and a quite strong magnetoresistance in LSMO/LNO/SRO trilayers using ferromagnetic resonance (FMR) at room temperature. Here, we demonstrate the influence of a LNO interlayer between LSMO and SRO on the effective spin mixing conductance of the different interfaces. The spin mixing conductance $g^{\uparrow\downarrow}$ is one of the key concepts in the spin current transport through interfaces\,\cite{Tserkovnyak} since it describes the transport of spins at the interface between a ferromagnet and a second layer made from a different material. It should be noted that according to Tserkovnyak \textit {et al.}\,\cite{Tserkovnyak} $g^{\uparrow\downarrow}$ only includes the transmission of said interface which is only a valid approach if the second layer exhibits very strong spin scattering. For a full description additional layer properties need to be taken into account, which is typically done using an effective spin mixing conductance $g^{\uparrow\downarrow}_\text{eff}$ instead of $g^{\uparrow\downarrow}$.  A large $g^{\uparrow\downarrow}_\text{eff}$ means a large spin current and if the difference in chemical potential of spin-up and spin-down in the ferromagnet is caused by ferromagnetic resonance, as in spin pumping experiments, it also means a larger damping of the resonance. However, the estimation of the spin mixing conductance in spin pumping experiments for magnetic/non-magnetic layer systems is not trivial. Usually the calculation is done by measuring the increase in damping and comparing it to the characteristic value of a single uncapped magnetic layer without the spin sink. This uncapped layer acts as a reference sample with no losses due to spin pumping.  However, in most of the metallic magnetic layers a capping layer is needed. The capping can largely modify the damping properties of the magnetic layer, a fact that cannot be correlated to the investigated spin pumping. Furthermore, different factors can influence the estimation of the increase in damping like, for example the emergence of a finite magnetic polarization in the non-magnetic layer (NM) in contact with a ferromagnetic layer\,\cite{Conca17,Caminale16}, the spin memory loss effect\,\cite{Rojas} or the two-magnon scattering effect\,\cite{Conca18}. This is the reason why in experiments typically $g^{\uparrow\downarrow}_\text{eff}$ is determined.
In our study no capping layer is needed because the bare LSMO reference layer is stable in air. Our work focuses on the estimation of $g^{\uparrow\downarrow}_\text{eff}$ in an oxide trilayer system. Previous experiments in which spin pumping through oxide interlayers was investigated showed different results. For some oxides\,\cite{Du2013} the spin pumping efficiency was reduced while for NiO\,\cite{Wang} the spin pumping efficiency and the inverse spin-Hall effect were increased. In our experiments we observe that the presence of an LNO interlayer increases the damping and as a consequence also $g^{\uparrow\downarrow}_\text{eff}$.  We derive the spin diffusion length for the interlayer LNO, as well as the effective spin mixing conductance for our trilayers.  Magnetization and Curie-temperatures are measured by SQUID magnetometry and the samples are structurally characterized by X-ray diffraction and transmission electron microscopy to confirm the interface quality.

\section{SAMPLE FABRICATION}
The heterostructures are deposited on (001)-oriented strontium titanate SrTiO$_3$ (STO) substrates, which are TiO$_2$-terminated by wet etching and annealing\,\cite{Schrott}. The deposition is done in a copper-sealed PLD chamber with a background pressure lower than 4$\cdot$10$^{-8}$\,mbar. For deposition an excimer laser with a wavelength of 248\,nm is used. The laser fluency is chosen as 2.3\,J\,cm$^{-2}$ per pulse and the repetition rate as 2\,Hz. An oxygen partial pressure of 0.2\,mbar for the LSMO (40\,nm) and SRO (6\,nm) layers, and 0.4\,mbar for the LNO layers is applied. Deposition is done at a substrate temperature of 750\,$^\circ$C. After deposition the heterostructure is cooled down at 5\,K\,min$^{-1}$. The heterostructure compositions and the thickness of each layer are summarized in Table\,\ref{Tabelle1}. For a LSMO/Pt reference stack (sample\,R2) a 13\,nm thick Pt layer is deposited via DC magnetron sputtering without breaking the vacuum. The samples are cut afterwards into pieces with the size of 2\,$\times$\,5\,mm$^2$ to fit into the SQUID magnetometer and the sample holder for FMR spectroscopy investigation.\\
For a better contact for ISHE measurements we deposited Ti(10\,nm)/Au(150\,nm), via lift-off process, on the edges of the respective samples.
\begin{table}
  \begin{tabular}{lcccc}
\hline
Name & LSMO  & LNO & SRO & Pt \\
 & [nm] & [nm] & [nm] & [nm]\\
\hline
R1 & 40 & & & \\
R2 & 40 & & & 13  \\
R3 & 40 & 3 & &\\
\hline
S1 & 40 & 0 & 6 & \\
S2 & 40 & 1.8 &6  &\\
S3 & 40 &3 &6  &\\
S4 & 40 & 6 & 6 & \\
S5 & 40 & 9&6 & \\
S6 & 40 & 11&6 & \\
S7 & 40 & 23&6 & \\
\hline
  \end{tabular}
\caption{Prepared layer stacks. The given thicknesses are nominal values.}\label{Tabelle1}
\end{table}

\section{STRUCTURAL CHARACTERIZATION}
For all samples structural characterization is done by X-ray diffraction and reflectometry. For sample\,S3, also high resolution transmission electron microscopy (HRTEM) was performed.
\begin{figure}[H]%
\centering
\includegraphics*[scale=0.47]{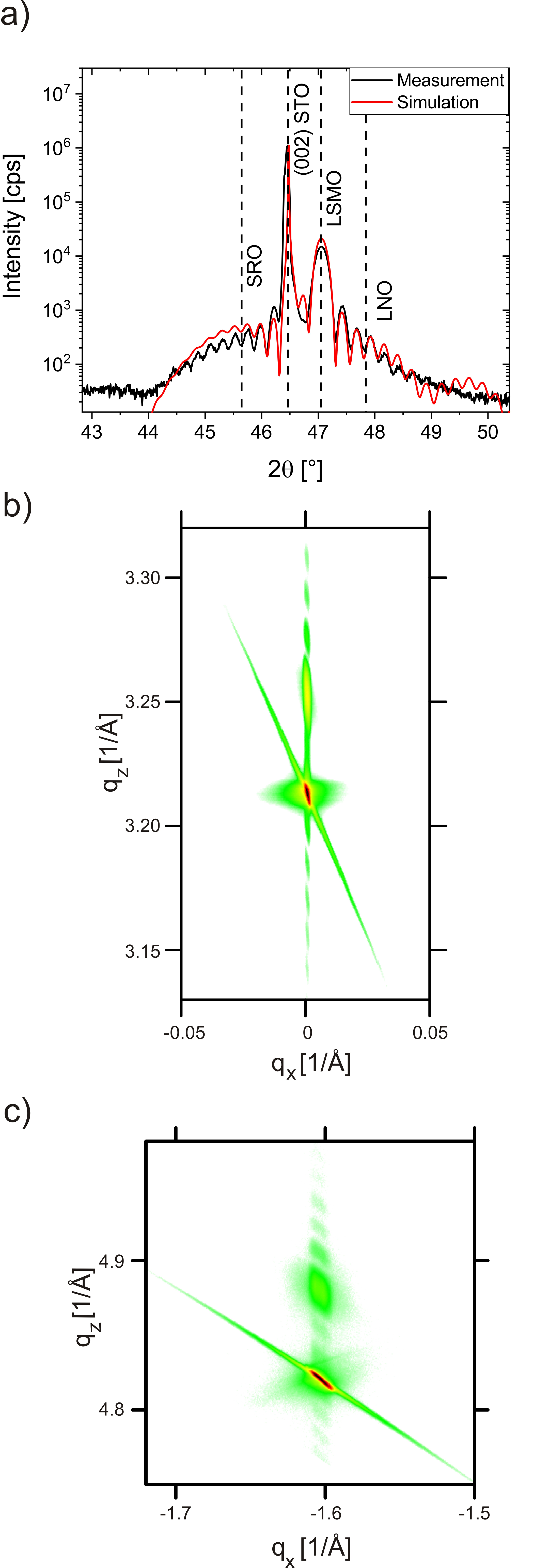}
\caption{a) $\omega$-2$\theta$ scan for the heterostructure  LSMO(40\,nm)/LNO(3\,nm)/SRO(6\,nm) (sample\,S3). The red curve is a fit to the data. b) and c) Reciprocal space maps for sample\,S3 of the symmetric (002) and the asymmetric ($\bar{1}$03) peak, respectively. The heterostructure is fully strained.}
\label{XRD}
\end{figure}
For X-ray characterization we use a Bruker D8 diffractometer with focussed CuK$_{\alpha1}$ radiation. For the detection a scintillation detector is used in an unlocked $\omega$-2$\Theta$ scan of the (002)\,-\,reflection. Figure\,\ref{XRD}\,a) shows, as an example, the results for sample\,S3. Simulations for fully strained layers yield out-of-plane lattice constants of c$_{\text{LSMO}}$\,=\,3.858\,\AA , c$_{\text{LNO}}$\,=\,3.730\,\AA\,\, and c$_{\text{SRO}}$\,=\,3.972\,\AA\,\, for our layer stacks. The dashed lines in  Figure\,\ref{XRD}\,a) mark the positions of the respective peaks for fully pseudomorphic materials as a guide to the eye. According to these results, diffraction peaks on the left hand side of the STO substrate (002) peak (cubic lattice constant\,\cite{Guo3} a$_{\text{STO,bulk}}$\,=\,3.905\,\AA) must stem from  SRO. On the right hand side the diffraction peaks of the layers with smaller lattice constant, namely LSMO and LNO appear. Due to the thinness of the LNO and SRO layers, only a prominent peak for the LSMO layer is visible. The presence of thickness fringes indicates smooth interfaces. A simulation with thicknesses of d$_\text{LSMO}$\,=\,39.1\,nm, d$_\text{LNO}$\,=\,2.5\,nm and d$_\text{SRO}$\,=\,5.5\,nm fits best the $\omega$-2$\Theta$ scan. The roughness of the surface and the interfaces calculated from X-ray reflectometry measurements is below 0.3\,nm for all heterostructures. In Figure\,\ref{XRD}\,b) and c) reciprocal space maps around the symmetric (002) and asymmetric ($\bar{1}$03) reflections are shown. As all layers have the same value for q$_\text{x}$ in the measurement around the asymmetric ($\bar{1}$03) peak, we can state that the heterostructure is fully strained and that the in-plane lattice constant for LSMO, LNO and SRO is equal to a$_{\text{STO}}$\,=\,3.905\,\AA.  The $\omega$-2$\Theta$ scan (Figure\,\ref{XRD}\,a)) is the line scan along q$_\text{x}=0$ for the measurement around the symmetric peak (Figure\,\ref{XRD}\,b)).
\begin{figure}[h]
\includegraphics*[scale=0.52]{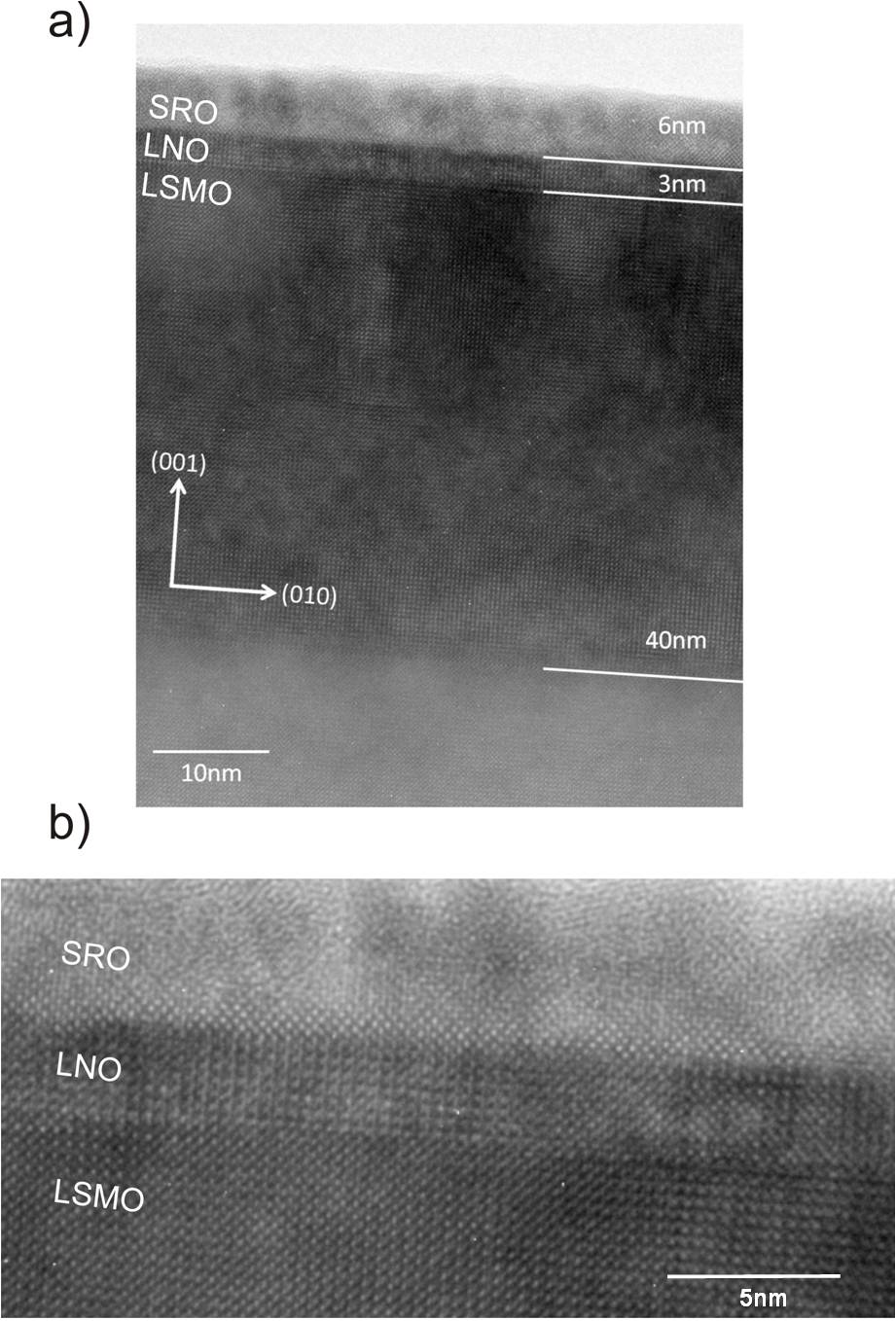}
\caption{HRTEM measurement for sample\,S3 STO/LSMO(40\,nm)/LNO(3\,nm)/SRO(6\,nm). The interfaces are smooth and LSMO and LNO appear monocrystalline while for SRO several crystal orientations can be seen.}
\label{TEM}
\end{figure}
For sample\,S3 a HRTEM image is made with a JEOL JEM 4010 electron microscope at an acceleration voltage of 400\,kV. Figure\,\ref{TEM}\,a) shows an image of the layer stack, which is cut in the (010) direction (imaging is done along the (100) direction). LSMO grows epitaxially with respect to the substrate and LNO is monocrystalline. For SRO the first monolayers are also epitaxial but apparently the rest of the layer is polycrystalline, as several orientations of the SRO crystal can be seen (Figure\,\ref{TEM}\,b)). The interfaces are smooth and a very low interface roughness ($<$\,0.4\,nm) is observed. The thicknesses shown in Figure\,\ref{TEM}\,a) match well the nominal thicknesses and the determined values from X-ray diffraction. We can thus estimate a deviation of less than 0.5\,nm from the nominal thicknesses listed in Table\,\ref{Tabelle1}.

\section{SQUID MAGNETOMETRY}
\begin{figure}[h]%
\begin{center}
\includegraphics*[scale=0.6]{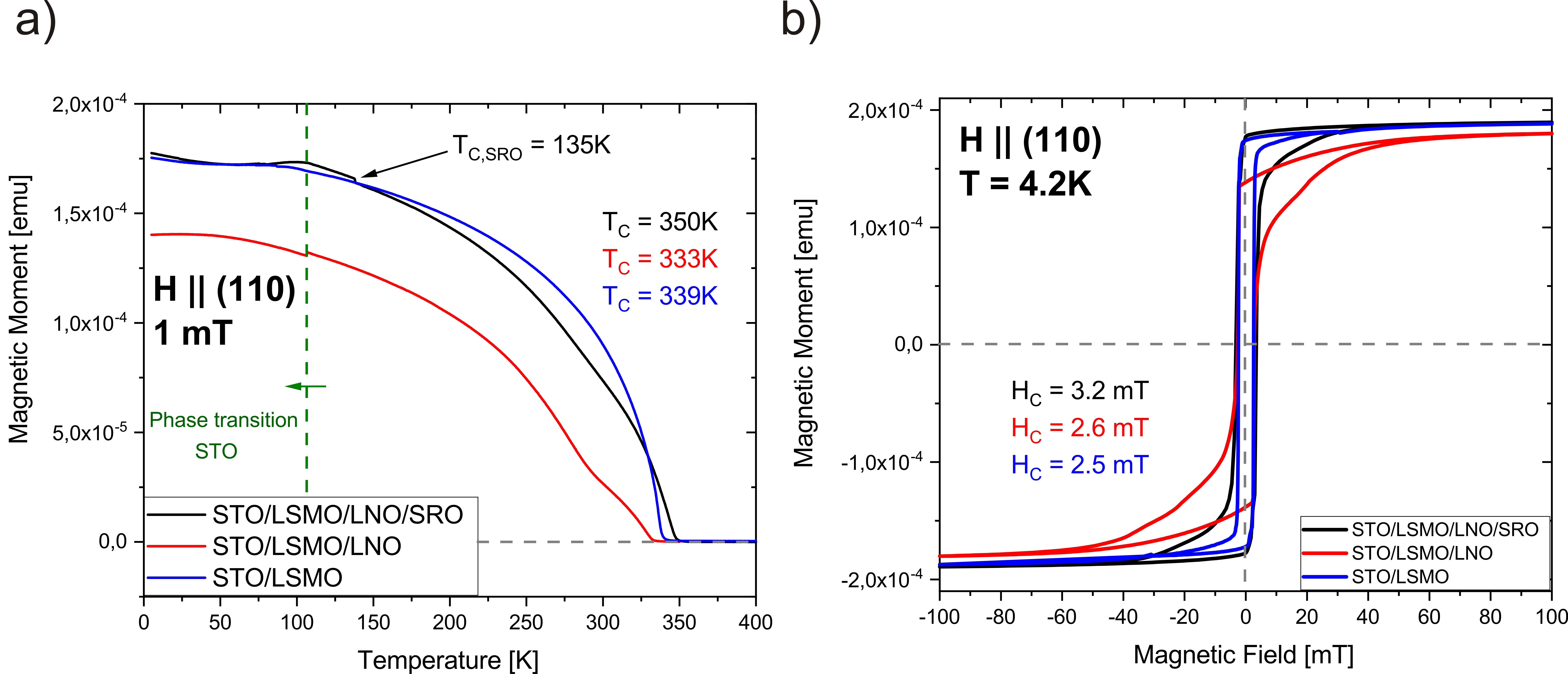}
\caption{a) Cooling curves for a bare LSMO (sample\,R1 blue), a LSMO/LNO (sample\,R3 red) and LSMO/LNO/SRO (sample\,S3 black) heterostructure, respectively.  The kink at 135\,K likely relates to the phase transition of the SRO to the ferromagnetic state. b) Hysteresis loops of the three samples. The coercive fields are symmetric around zero for all samples}
\label{SQUID}
\end{center}
\end{figure}

The magnetic characterization of samples\,R1, R3, and S3 is done with a Quantum Design SQUID VSM magnetometer. Here we measure cooling curves and hysteresis loops for an LSMO/LNO (R3) layer stack and compare the results to the measurements for a bare LSMO layer (R1). For the investigation of the ferromagnetic phase transition of SRO we additionally measure an LSMO/LNO/SRO  heterostructure (S3).  The cooling curves (Figure\,\ref{SQUID}\,a)) are measured at an external field of 1\,mT along the (110) direction of LSMO, which is the magnetic easy axis (according to our FMR experiments). The Curie temperature of LSMO for all samples is between 333\,-\,350\,K, which is slightly lower than the value of 370\,K for bulk material\,\cite{Sadoc}. The drop of the magnetization below 105\,K for sample\,S3 and R1 as well as the kink in the cooling curve of sample\,R3 at 105\,K can be explained with the structural phase transition from cubic to tetragonal structure of the STO substrate at T\,$\approx$\,105\,K\,\cite{Loetzsch,Cowley,Lytle}. Angular dependence of the resonance field suggests a change in magnetic anisotropy of LSMO due to the phase transition of the STO (not shown here), which hence leads to a change in the measured magnetic moment. The kink at 135\,K for sample\,S3 likely marks the ferromagnetic phase transition of SRO (T$_C$\,=\,155\,K for bulk material\,\cite{Koster,Wahler}). In Figure\,\ref{SQUID}\,b) hysteresis loops for samples\,R1, R3 and S3 are shown. The measurements are done with zero-field cooling, as this is also the way the FMR experiments are done. Here the respective coercive fields for all three layer stacks are comparable and we do not observe any exchange bias in contrast to Guo \textit{et al.}\,\cite{Guo} who found that bulk LNO becomes antiferromagnetic below 157\,K if fabricated under certain deposition conditions and Sanchez \textit{et al.}\,\cite{Sanchez} and Peng \textit{et al.}\,\cite{Peng} who reported an exchange bias in LSMO/LNO bilayers below 50\,K and 100\,K, respectively.

\section{FERROMAGNETIC RESONANCE AND SPIN PUMPING}
\begin{figure}[h]%
\includegraphics*[scale=0.52]{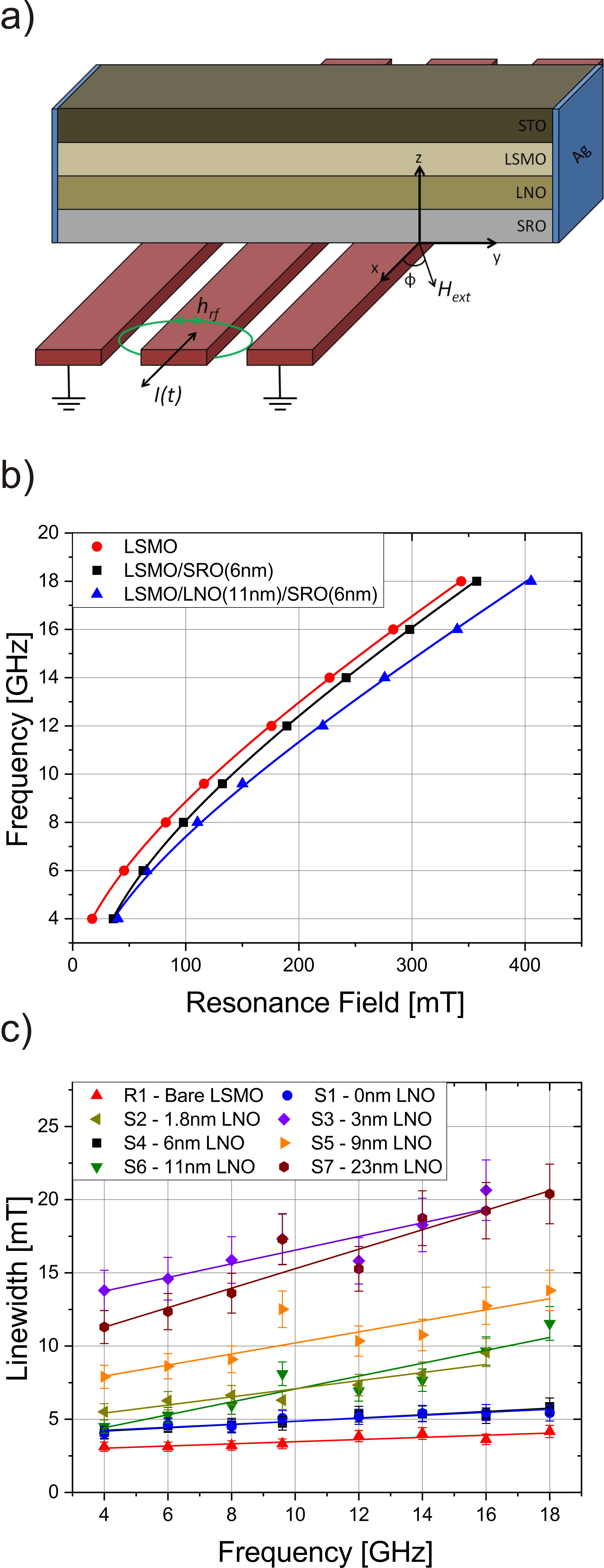}
\caption{a) Measurement geometry for the FMR investigation. b) Resonance frequency in dependence of the resonance field. The lines are fits to Kittel's equation. c) FMR linewidth in dependence of the resonance frequency. The lines are linear fits to estimate the damping parameter $\alpha$.}
\label{FMR-K-and-D}
\end{figure}

\begin{figure}[h]%
\includegraphics*[scale=0.52]{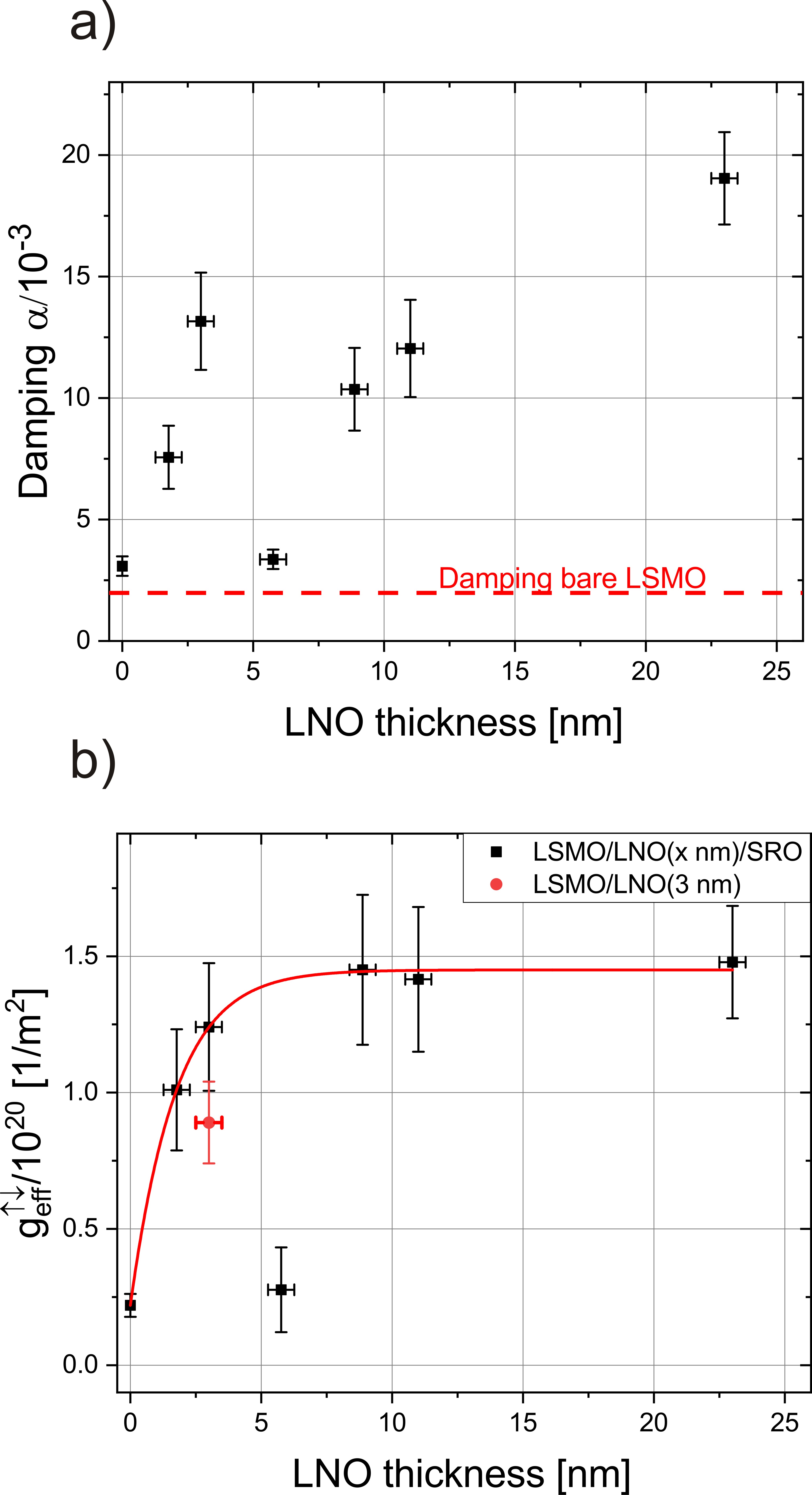}
\caption{a) Damping and b) effective spin mixing conductance in dependence of the interlayer thickness in LSMO/LNO/SRO trilayers. The red line in b) is a fit to the shown data and equation\,\ref{dampTri}. The fit reveals a spin diffusion length of $\lambda_\text{LNO}$\,=\,3.3\,nm.}
\label{FMR}
\end{figure}

\begin{figure}[htb]%
\begin{center}
\includegraphics*[scale=0.45]{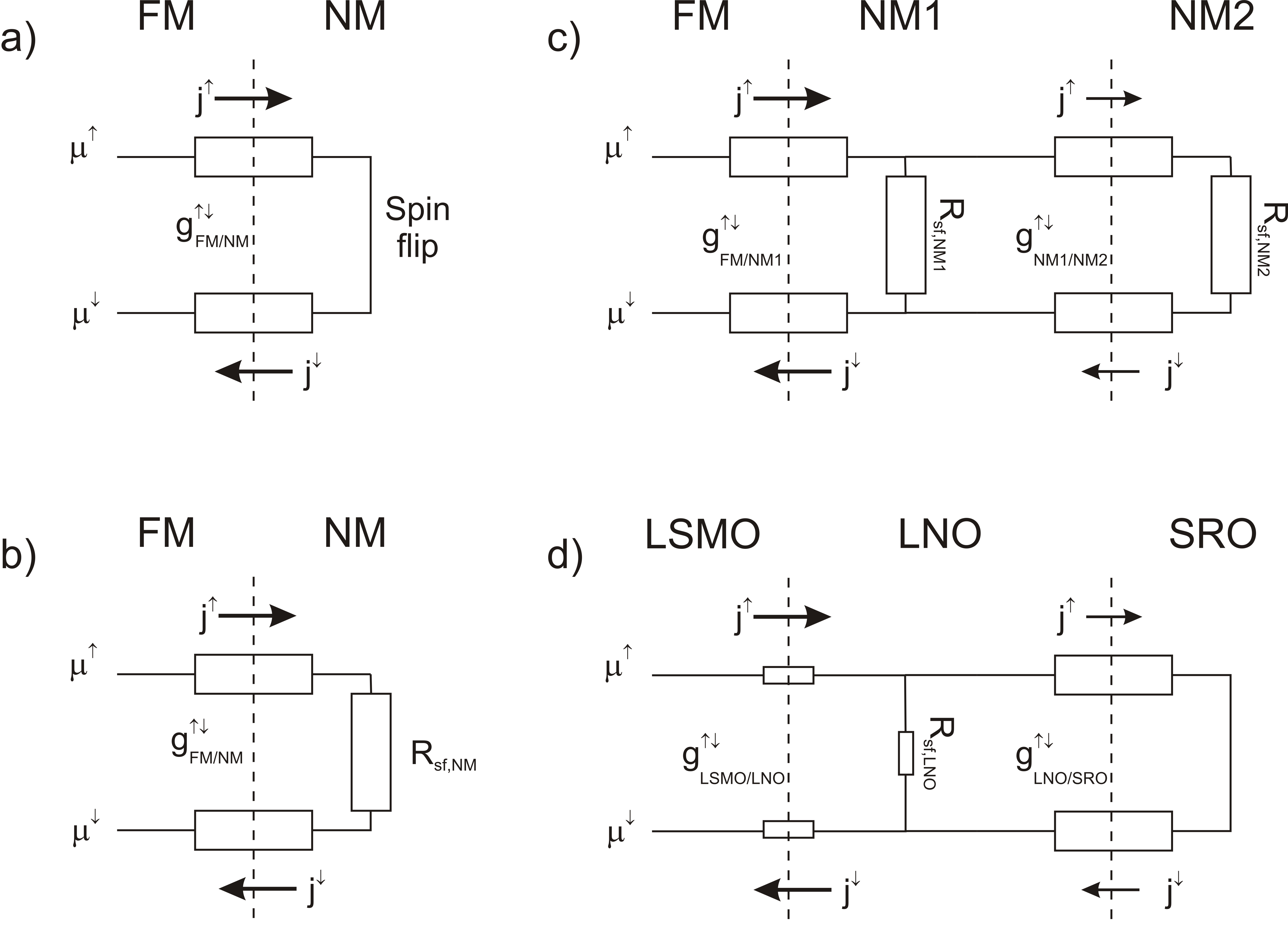}
\caption{Schematic diagram of the effective spin mixing conductance. Due to the FMR a spin accumulation is formed at the interface FM/NM which results in a splitting of the chemical potential $\mu$ of $\uparrow$- and $\downarrow$-states a) Model circuit for the effective spin mixing conductance in a bilayer system FM/NM in case of immidiate spin flip in the non-magnet. b) For finite spin diffusion length and resistance in NM a spin flip resistance $R_\text{sf,NM}$ has to be added. c) For trilayers FM/NM1/NM2 a second interface is generated with an additional spin sink NM2, which can be described using the interface transmission $g_\text{NM1/NM2}^{\uparrow\downarrow}$ and a second spin flip resistance $R_\text{sf,NM2}$. d) In our LSMO/LNO/SRO trilayer $g_\text{LSMO/LNO}^{\uparrow\downarrow}$ and $1/R_\text{sf,LNO}$ are both much larger than $g_\text{LNO/SRO}^{\uparrow\downarrow}$. This is pointed out by the size of the corresponding resistors (large $g$\,$\rightarrow$\,small $R$).}
\label{SMC}
\end{center}
\end{figure}

FMR measurements and the investigation of the ISHE are performed in a cryostat at 190\,K which, as shown previously, is below the T$_C$ of LSMO and above the T$_C$ of SRO, in order to achieve spin pumping from a ferromagnetic into two paramagnetic layers. In addition, the choice of 190\,K prevents a vanishing ISHE in SRO as was observed at lower temperature by Wahler \textit{et al.}\,\cite{Wahler}. A RF current through a coplanar waveguide generates the necessary excitation field for FMR. To saturate the sample magnetization a rotatable electromagnet generates a homogeneous external magnetic field aligned along the waveguide and along the (010) direction of LSMO (in-plane, $\phi\,=\,0^\circ$), which is shown in Figure\,\ref{FMR-K-and-D}\,a).
For measuring FMR, a continuous-wave signal at constant frequency $f$ in the range of 4\,-\,18\,GHz is applied at a power of 10\,dBm. The RF transmission is measured using a diode. The external magnetic field is modulated with 0.2\,mT amplitude at a frequency of 20\,Hz  and lock-in technique is used to improve the signal-to-noise ratio.\\
The FMR field-swept measurements are fitted with a derivative of a Lorentzian function, which yields the half-width at half maximum FMR linewidth $\triangle H_\text{FMR}$ and resonance field $H_\text{FMR}$\,\cite{Celinski}. With Kittel's equation\,\cite{Liu2}:
\begin{align}\label{Kittel}
f=\dfrac{\mu_0\gamma}{2\pi}\sqrt{(H_{\text{res}}+H_{\text{ani}})(H_{\text{res}}+H_{\text{ani}}+M_\text{eff})}
\end{align}
we derive the gyromagnetic ratio $\gamma$ and the effective magnetization $M_\text{eff}$ from the  dependence between $f$ and the resonance field $H_\text{res}$. $\mu_0$ and $H_\text{ani}$ are the vacuum permeability and the anisotropy field, respectively.  For $\gamma /2\pi$ and $\mu_0 H_\text{ani}$ we derive values in the range of (27.5\,-\,28.5)\,GHz\,T$^{-1}$ and (2.4\,-\,33.9)\,mT, respectively. Indicatively, the results for samples\,R1, S1 and S6 are shown in Figure\,\ref{FMR-K-and-D}\,b). The damping parameter $\alpha$ can be derived from a linear fit of the FMR linewidth $\triangle H_\text{FMR}$ plotted over the RF-frequency\,\cite{Liu3}\,$f$:
\begin{align}
\mu_0\triangle H_{\text{FMR}}=\dfrac{2\pi\alpha f}{\gamma}+\mu_0\triangle H_0
\end{align}
where $\triangle H_0$ is the inhomogeneous linewidth broadening. The values of $\alpha$, $\triangle H_0$, and $M_\text{eff}$ for all our heterostructures are summarized in Table\,\ref{Tabelle2} and damping measurements for samples\,R1 and S1\,-\,S7 are shown in Figure\,\ref{FMR-K-and-D}\,c). In our experiments we see that the bare LSMO film (sample\,R1) exhibits the lowest damping ($\alpha_\text{LSMO}$\,=\,2.0\,$\cdot$\,10$^{-3}$). With any layer on top of LSMO that can act as a spin sink, the damping increases significantly, showing that the spin sink layers introduce additional channels for  loss of angular momentum. Figure\,\ref{FMR}\,a) shows the dependence of the damping as a function of the LNO interlayer thickness for samples S1\,-\,S7. The increase of damping in the heterostructures can be understood as the generated spin current carries angular momentum from the ferromagnet (FM) into the nonmagnetic interlayer (NM1) where it is either lost to the lattice by spin flip or diffuses further into the final spin sink (NM2). Due to conservation of angular momentum a torque will force the magnetization to decrease the precession angle and with this increase the damping of the FM. But no clear trend of the damping in dependence of the LNO interlayer thickness can be observed here.
Only when we do a full calculation of the effective spin mixing conductance from equation\,(\ref{spinmix}) using the extracted $M_\text{s}\approx M_\text{eff}$ from equation\,(\ref{Kittel}) and $\alpha_\text{sp} = \alpha_\text{sample}-\alpha_\text{LSMO}$ the dependence becomes visible. The derived values of $g^{\uparrow\downarrow}_\text{eff}$ in the LSMO/LNO/SRO trilayers are depicted as a function of the LNO thickness in Figure\,\ref{FMR}\,b), showing a clear increase with LNO thickness and saturating at $\approx$\,1.45\,$\cdot$\,10$^{20}$\,m$^{-2}$ for approx. 9\,nm LNO interlayer thickness.
In order to interpret our following experiments it is necessary to understand the effective spin mixing conductance in a trilayer system which has several contributions that may even influence each other. The spin mixing conductance $g^{\uparrow\downarrow}$ introduced by Brataas \textit {et al.}\,\cite{Brataas} and Tserkovnyak \textit{et al.}\,\cite{Tserkovnyak} only quantifies the spin transmission through the interface between the ferromagnet (FM) and the adjacent non-magnet (NM). To simplify a further discussion we rename $g^{\uparrow \downarrow}$ to $g^{\uparrow \downarrow}_\text{FM/NM}$. The quantity $g^{\uparrow\downarrow}_\text{FM/NM}$ does not take into account properties of the non-magnet like conductivity or spin diffusion length. Only in the case where spins entering the NM layer are immediately flipped, this spin mixing conductance alone needs to be considered for the additional damping by spin pumping as done by Tserkovnyak \textit{et al.}\,\cite{Tserkovnyak}. Starting from a bilayer system with immediate spin flip we consinder a model for the spin mixing conductance which is shown in Figure\,\ref{SMC}\,a). As soon as a spin accumulation appears in the non-magnet the spin flow through the spin mixing conductance is reduced and the spin current is no longer defined by $g^{\uparrow\downarrow}_\text{FM/NM}$ but by $g^{\uparrow\downarrow}_\text{eff}$. In an equivalent circuit (Figure\,\ref{SMC}\,b)) this can be implemented by adding a resistance $R_\text{sf,NM}$ between the spin accumulation ($\mu^\uparrow$ and $\mu^\downarrow$) in the non-magnet which represents the spin flip, necessary to accomodate the steady state of one spin direction flowing into the non-magnet and the other flowing back. Only for immediate and complete spin flip this resistance is a short circuit leading back to $g^{\uparrow\downarrow}_\text{eff}=g^{\uparrow\downarrow}_\text{FM/NM}$. For a finite resistivity $\sigma_\text{NM}$ and finite spin diffusion length $\lambda_\text{NM}$ the effective spin mixing conductance can be calculated from the additional damping in a bilayer system\,\cite{Mosendz3,Du} for $\lambda_\text{NM}\ll d_\text{NM}$:
\begin{align}\label{spinmix}
\alpha_\text{sp}&=\dfrac{g_\text{eff}^{\uparrow\downarrow}g_\text{L}\mu_\text{B}}{4\pi d_\text{FM} M_\text{s}}\\
\text{with:}\,\,\,\dfrac{1}{g_\text{eff}^{\uparrow\downarrow}}&=\dfrac{1}{g^{\uparrow\downarrow}_\text{FM/NM}}+R_\text{sf,NM}\nonumber
\end{align}
($g_\text{L}$:  g-factor, $\mu_\text{B}$: Bohr magneton and $M_\text{s}$: saturation magnetization, $d$: layer thickness)
where the magnitude of the resistance $R_\text{sf,NM}$ depends on the spin flip time $\tau_\text{SF}$ of the NM  ($\lambda_\text{NM}\propto\sqrt{\tau_\text{SF}}$) and only for a thickness of the NM ($d_\text{NM}$) much bigger than $\lambda_\text{NM}$ it has the fixed value of $R_\text{sf,NM}\propto\lambda_\text{NM}/\sigma_\text{NM}$. For a thinner NM film the so called back flow needs to be taken into account and the more complex expression\,\cite{Tserkovnyak}
\begin{align}\label{addDamp}
\alpha_\text{sp}=&\Bigg[ 1+ g^{\uparrow\downarrow}_\text{FM/NM}\dfrac{\tau_\text{SF}\delta_\text{SD}}{h\tanh (d_\text{NM}/\lambda_\text{NM})}\Bigg]^{-1}\cdot\nonumber\\ &\cdot\dfrac{g_Lg^{\uparrow\downarrow}_\text{FM/NM}\mu_\text{B}}{4\pi d_\text{FM} M_\text{s}}
\end{align}
for the additional damping in a bilayer system needs to be used. Here $\delta_\text{SD}$ is the energy level between two scattering states. Equation\,\ref{spinmix} is the limiting case of equation\,\ref{addDamp} for large $d_\text{NM}$. In our experiments the limit of $\lambda_\text{NM}\ll d_\text{NM}$ is not yet reached and any addition of another layer will further increase $g^{\uparrow\downarrow}_\text{eff}$. In case a third layer (NM2) is added, the interface between NM1 and NM2 needs to be considered again in a similar way as for the first interface now adding the spin transmission $g^{\uparrow\downarrow}_\text{NM1/NM2}$ to our picture (Figure\,\ref{SMC}\,c)). And again we have to take into account the layer properties of NM2 by adding $R_\text{sf,NM2}$. The resulting additional damping $\alpha_\text{sp}$ in a trilayer system can according to Tserkovnyak \textit{et al.}\,\cite{Tserkovnyak} be written as:
\begin{align}\label{dampTri}
&\alpha_\text{sp}=\Bigg[ 1+g_\text{FM/NM1}^{\uparrow\downarrow}\dfrac{\tau_\text{SF}\delta_\text{SD}}{h}\cdot\nonumber\\&\cdot\dfrac{1+\tanh(d_\text{NM1}/\lambda_\text{NM1})\tau_\text{SF}\delta_\text{SD}g^{\uparrow\downarrow}_\text{NM1/NM2}/h}{\tanh(d_\text{NM1}/\lambda_\text{NM1})+\tau_\text{SF}\delta_\text{SD}g^{\uparrow\downarrow}_\text{NM1/NM2}/h} \Bigg]^{-1}\cdot\nonumber\\&\cdot\dfrac{g_\text{L}g^{\uparrow\downarrow}_\text{FM/NM1}\mu_\text{B}}{4\pi d_\text{FM} M_\text{s}}
\end{align}
It should be noted that also in this equation Tserkovnyak \textit{et al.}\,\cite{Tserkovnyak} assume immediate spin flip in the third layer so that for the third layer only the transmission through the interface to the second one needs to be considered.
For the limiting case of $\lambda_\text{NM1}\ll d_\text{NM1}$ the third layer should have no influence any more and indeed we find that in this limit the result of equation\,\ref{dampTri} becomes identical to that of equation\,\ref{addDamp} only with $\lambda_\text{NM1}\ll d_\text{NM1}$ replaced by $\lambda_\text{NM}\ll d_\text{NM}$.\\
It should be noted that the introduced model (Figure\,\ref{SMC}) does not include a spin flip at the interface by scattering which would be associated with the so called spin memory loss\,\cite{Rojas}. The equivalent circuit might be extended to include this effect but this will be described elsewhere.\\
For our measurements of the effective spin mixing conductance, shown in Figure\,\ref{FMR}\,b), we can fit the data to equation\,\ref{dampTri}. SRO has a low spin diffusion length so the assumption of immediate spin flip after entering the SRO is valid and only $g_\text{NM1/NM2}^{\uparrow\downarrow}$ for the LNO/SRO interface, $g_\text{FM/NM1}^{\uparrow\downarrow}$ for the LSMO/LNO interface and the spin flip and resistance of the LNO need to be taken into account. Here we consider $g_\text{eff}^{\uparrow\downarrow}\propto \alpha_\text{sp}$. In most of the published experiments the interlayer exhibits little spin flip while the spin sink (e.g. Pt)  has a very high spin mixing conductance. In this case we have $g_\text{NM1/NM2}^{\uparrow\downarrow}>\frac{h}{\tau_\text{SF}\delta_\text{SD}}$ and an increase in thickness of the interlayer results in a decrease of the effective spin mixing conductance, as for example in FM/NM1/Pt trilayer systems\,\cite{Tserkovnyak,Deorani,Du}. For LNO apparently the spin diffusion length is small, which combined with a large conductivity leads to $g_\text{NM1/NM2}^{\uparrow\downarrow}<\frac{h}{\tau_\text{SF}\delta_\text{SD}}$ and we get an increase of the spin mixing conductance with increasing interlayer thickness. We can fit the data with a spin diffusion length of $\lambda_\text{LNO}=(3.3\pm 0.9)$\,nm for LNO. Taking the error bars of the measured data into account  we can set a lower limit for $\lambda_\text{LNO}$ at 1.7\,nm.\\
We can now compare the different contributions  to the effective SMC. When a 6\,nm SRO layer is put on LSMO (sample\,R1\,$\rightarrow$\,sample\,S1) the effective spin mixing conductance increases from 0 to $2\cdot 10^{19}$\,m$^{-2}$. Adding 3\,nm of LNO onto LSMO (sample\,R1\,$\rightarrow$\,sample\,R3) increases the effective spin mixing conductance from 0 to $~9\cdot 10^{19}$\,m$^{-2}$. Thus we assume that $g^{\uparrow\downarrow}$ for LSMO/LNO is bigger than for LSMO/SRO. The ratio must even be more than 9:2 because we found that for 3\,nm of LNO the spins are not yet flipped completely but some backflow occurs. When SRO is added to the LSMO/LNO bilayer (sample\,R3\,$\rightarrow$\,sample\,S3) the increase of $g_\text{eff}^{\uparrow\downarrow}$ is even identical within the error bars to the transition from pure LSMO to LSMO/SRO (sample\,R1\,$\rightarrow$\,sample\,S1). This leads to the following picture for all parts of $g_\text{eff}^{\uparrow\downarrow}$: The interface contribution  $g_\text{LSMO/LNO}^{\uparrow\downarrow}$ is much larger than $g_\text{LSMO/SRO}^{\uparrow\downarrow}$ for the LSMO/SRO bilayer. $g_\text{LNO/SRO}^{\uparrow\downarrow}$ is similar to $g_\text{LSMO/SRO}^{\uparrow\downarrow}$. Because of the extremely short spin diffusion length in SRO we can consider the connection between spin-up and spin-down channel in SRO as a short circuit, consistent with equation\,\ref{dampTri}. The spin diffusion length in LNO is comparable to the layer thickness so the spin-flip conductance $1/R_\text{sf,LNO}$ has a finite value. However, from figure\,\ref{SMC}\,b) it becomes clear, that just because of Ohm's law, the values of  $1/R_\text{sf,LNO}$ and $g_\text{LSMO/LNO}^{\uparrow\downarrow}$ both must be larger than $g_\text{eff}^{\uparrow\downarrow}$ of the LSMO/LNO bilayer (because all three resistors are in series)  and hence are also much larger than $g_\text{LNO/SRO}^{\uparrow\downarrow}$. In Figure\,\ref{SMC}\,d) this is depicted by the size of the different resistors (large $g$\,$\rightarrow$\,small $R$). \\
It is important to understand that the increase in $g_\text{eff}^{\uparrow\downarrow}$ when a LNO interlayer between LSMO and SRO is introduced is mainly due to the spin-flip and the large conductivity of LNO. Even if the transmission through the LSMO/SRO interface were perfect ($g_\text{LSMO/LNO}^{\uparrow\downarrow}\rightarrow\infty$) the insertion of the LNO layer would not increase $g_\text{eff}^{\uparrow\downarrow}$ but mainly leave it constant because we know that $g_\text{LNO/SRO}^{\uparrow\downarrow}\approx g_\text{LSMO/SRO}^{\uparrow\downarrow}$. The increase only can occur if an additional spin flip channel is created inside the LNO layer. It should be noted that also spin memory loss at the LSMO/LNO interface might be a cause but the evident dependence of $g_\text{eff}^{\uparrow\downarrow}$ on the LNO thickness tells us otherwise.
Our maximum values for the effective spin mixing conductance are higher than the recently published values of Ghosh \textit{et al.}\,\cite{Ghosh}, who estimated the effective spin mixing conductance in LSMO/LNO/SRO trilayers at room temperature. Most likely this is due to the increase in conductivity of the samples at lower temperature which increases $g_\text{sf,LNO}^{\uparrow\downarrow}$.

Because $g^{\uparrow\downarrow}$ is related to the Sharvin resistance\,\cite{Deorani} it is also understandable that its value increases with conductivity of the spin sink and the number of available conducting channels. This assumption is well in line with our results for sample\,R2 (LSMO/Pt) which has the highest effective spin mixing conductance of all samples and the highest conductivity with a pure metal spin sink.
\begin{table}
\begin{center}
  \begin{tabular}{lccc}
\hline
Sample &  Damping $\alpha/10^{-3}$ & $M_\text{eff}/10^5$ &  $g_\text{eff}^{\uparrow\downarrow}/10^{20}$\\
 &  & [A/m] &  [1/m$^2$]\\
\hline
R1 & 2.0\,$\pm$\,0.1 & 6.8\,$\pm$\,0.1 & -\\
R2 &  20.0\,$\pm$\,3.0 &6.9\,$\pm$\,0.3 &  3.31\,$\pm$\,0.53\\
R3 &  11.0\,$\pm$\,1.0 & 3.8\,$\pm$\,0.2 &  0.89\,$\pm$\,0.15\\
\hline
S1 &  3.1\,$\pm$\,0.4 & 7.5\,$\pm$\,0.2 & 0.22\,$\pm$\,0.04\\
S2 &  7.6\,$\pm$\,1.3 & 6.7\,$\pm$\,0.3 &  1.01\,$\pm$\,0.22\\
S3 &  13.0\,$\pm$\,2.0 & 4.1\,$\pm$\,0.2 &  1.24\,$\pm$\,0.24\\
S4 & 3.4\,$\pm$\,0.3 & 7.5\,$\pm$\,0.8 &  0.28\,$\pm$\,0.16\\
S5 &  10.4\,$\pm$\,1.7 & 6.4\,$\pm$\,0.1 &  1.45\,$\pm$\,0.28\\
S6 &  12.0\,$\pm$\,2.0 & 5.2\,$\pm$\,0.1 & 1.42\,$\pm$\,0.27\\
S7 & 19.0\,$\pm$\,1.9 & 3.2\,$\pm$\,0.1 &  1.48\,$\pm$\,0.21\\
\hline
  \end{tabular}
    \caption{Summarized results from FMR measurements}
 \label{Tabelle2}
  \end{center}
\end{table}

Finally, it should be noted that we also tried to measure the inverse spin-Hall effect (ISHE)\,\cite{Wei,Harii} in the different samples. We investigated the ISHE voltage for all fabricated samples, by separating it from the anisotropic magnetoresistance effect\,\cite{Wahler, Azevedo, Mecking, Obstbaum} and thermo voltages\,\cite{Ando2}. The results, however, are inconclusive. For LSMO/LNO, no ISHE can be detected within our measurement accuracy. Although for trilayer samples a small effect can be detected, the ISHE is much smaller than for LSMO/SRO and shows no clear dependence on the LNO thickness.

\section{Conclusion}
We have shown that the insertion of a LNO layer between LSMO and SRO increases the effective SMC. This effect can be linked to a highly transparent interface between LSMO and LNO and a large spin flip in the highly conducting LNO. Thickness dependent measurements indicate a spin diffusion length of approx. 3.3\,nm which is still twice as long as shown for SRO.\,\cite{Wahler} $g^{\uparrow\downarrow}$ for LSMO/SRO and for LNO/SRO seem to be of similar magnitude. The increase for the effective spin mixing conductance leads to increased damping, however, only the outflow of spin current from the LSMO but not the inflow of spin current into the SRO is increased.

\section{Acknowledgement}
This work was supported by the SFB\,762. We thank the Max-Planck-Institut for microstructure physics for the access to transmission electron microscopy.

\newpage

\begin{thebibliography}{10}

\bibitem{Sinova}
J.~Sinova, S.~O. Valenzuela, J.~Wunderlich, and C.~Back, ``Spin hall effects,''
  {\em Rev. Mod. Phys.}, vol.~87, p.~1213, 2015.

\bibitem{Hoffmann}
A.~Hoffmann, ``Spin hall effects in metals,'' {\em IEEE Transactions on
  Magnetics}, vol.~49, p.~5172, 2013.

\bibitem{Cui}
B.~Cui, C.~Song, G.~Wang, Y.~Yan, J.~Peng, J.~Miao, H.~Mao, F.~Li, C.~Chen,
  F.~Zeng, and F.~Pan, ``Reversible ferromagnetic phase transition in
  electrode-gated manganites,'' {\em Adv. Funct. Mater.}, vol.~24, p.~7233,
  2014.

\bibitem{Klein}
L.~Klein, J.~S. Dodge, C.~H. Ahn, G.~J. Snyder, T.~H. Geballe, M.~R. Beasley,
  and A.~Kapitulnik, ``Anomalous spin scattering effects in the badly metallic
  itinerant ferromagnet srruo$_3$,'' {\em Phys. Rev. Lett.}, vol.~77, p.~2774,
  1996.

\bibitem{Koster}
G.~Koster, L.~Klein, W.~Siemons, G.~Rijnders, J.~S. Dodge, C.~B. Eom, D.~H.~A.
  Blank, and M.~R. Beasley, ``Structure, physical properties, and applications
  of srruo$_3$ thin films,'' {\em Rev. Mod. Phys.}, vol.~84, p.~253, 2012.

\bibitem{Guo}
H.~Guo, Z.~W. Lie, L.~Zhao, Z.~Hu, C.~F. Chang, C.-Y. Kuo, W.~Schmidt,
  A.~Piovano, T.~W. Pi, O.~Sobolev, D.~I. Khomskii, L.~Tjeng, and A.~C.
  Komarek, ``Antiferromagnetic correlations in the metallic strongly correlated
  transition metal oxide lanio$_3$,'' {\em Nat. Commun.}, vol.~9, p.~43, 2018.

\bibitem{Wahler}
M.~Wahler, N.~Homonnay, T.~Richter, A.~M\"uller, B.~Fuhrmann, and G.~Schmidt,
  ``Inverse spin hall effect in a complex ferromagnetic oxide
  heterostructure,'' {\em Sci. Rep.}, vol.~6, p.~28727, 2016.

\bibitem{Ghosh}
S.~Ghosh, R.~G. Tanguturi, P.~Pramanik, D.~C. Joshi, P.~K. Mishra, S.~Das, and
  S.~Thota, ``Low-temperature anomalous spin correlations and kondo effect in
  ferromagnetic srruo$_3$/lanio$_3$/la$_{0.7}$sr$_{0.3}$mno$_3$ trilayers,''
  {\em Phys. Rev. B}, vol.~99, p.~115135, 2019.

\bibitem{Tserkovnyak}
Y.~Tserkovnyak, A.~Brataas, and G.~E.~W. Bauer, ``Spin pumping and
  magnetization dynamics in metallic multilayers,'' {\em Phys. Rev. B},
  vol.~66, p.~224403, 2002.

\bibitem{Conca17}
A.~Conca, B.~Heinz, M.~R. Schweizer, S.~Keller, E.~T. Papaioannou, and
  B.~Hillebrands, ``Lack of correlation between the spin-mixing conductance and
  the inverse spin hall effect generated voltages in cofeb/pt and cofeb/ta
  bilayers,'' {\em Phys. Rev. B}, vol.~95, p.~174426, 2017.

\bibitem{Caminale16}
M.~Caminale, A.~Ghosh, S.~Auffret, U.~Ebels, K.~Ollefs, F.~Wilhelm, A.~Rogalev,
  and W.~E. Bailey, ``Spin pumping damping and magnetic proximity effect in pd
  and pt spin-sink layers,'' {\em Phys. Rev. B}, vol.~94, p.~014414, 2016.

\bibitem{Rojas}
J.-C. Rojas-S\'{a}nchez, N.~Reyren, P.~Laczkowski, W.~Savero, J.-P.
  Attane\'{e}, C.~Deranlot, M.~Jamet, J.-M. George, L.~Villa, and
  H.~Jaffr\'{e}s, ``Spin pumping and inverse spin hall effect in platinum: The
  essential role of spin-memory loss at metallic interfaces,'' {\em Phys. Rev.
  Lett.}, vol.~112, p.~106602, 2014.

\bibitem{Conca18}
A.~Conca, S.~Keller, M.~R. Schweizer, E.~T. Papaioannou, and B.~Hillebrands,
  ``Separation of the two-magnon scattering contribution to damping for the
  determination of the spin mixing conductance,'' {\em Phys. Rev. B}, vol.~98,
  p.~214439, 2018.

\bibitem{Du2013}
C.~H. Du, H.~L. Wang, Y.~Pu, T.~L. Meyer, P.~M. Woodward, F.~Y. Yang, and P.~C.
  Hammel, ``Probing the spin pumping mechanism: Exchange coupling with
  exponential decay in
  ${\mathbf{y}}_{3}{\mathrm{fe}}_{5}{\mathbf{o}}_{12}/\mathrm{\text{barrier}}/\mathrm{Pt}$
  heterostructures,'' {\em Phys. Rev. Lett.}, vol.~111, p.~247202, 2013.

\bibitem{Wang}
H.~Wang, C.~Du, P.~C. Hammel, and F.~Yang, ``Antiferromagnonic spin transport
  from y$_3$fe$_5$o$_{12}$ into nio,'' {\em Phys. Rev. Lett.}, vol.~113,
  p.~097202, 2014.

\bibitem{Schrott}
A.~G. Schrott, J.~A. Misewich, M.~Copel, D.~W. Abraham, and Y.~Zhang, ``A-site
  surface termination in strontium titanate single crystals,'' {\em Appl. Phys.
  Lett.}, vol.~79, p.~1786, 2001.

\bibitem{Guo3}
H.~Guo, S.~Dong, P.~D. Rack, J.~D. Budai, C.~Beekman, Z.~Gai, W.~Siemons, C.~M.
  Gonzalez, R.~Timilsina, A.~T. Wong, A.~Herklotz, P.~C. Snijders, E.~Dagotto,
  and T.~Z. Ward, ``Strain doping: Reversible single-axis control of a complex
  oxide lattice via helium implantation,'' {\em Phys. Rev. Lett.}, vol.~114,
  p.~256801, 2015.

\bibitem{Sadoc}
A.~Sadoc, B.~Mercey, C.~Simon, D.~Grebille, W.~Prellier, and M.-B. Lepetit,
  ``Large increase of the curie temperature by orbital ordering control,'' {\em
  Phys. Rev. Lett.}, vol.~104, p.~046804, 2010.

\bibitem{Loetzsch}
R.~Loetzsch, A.~L\"ubcke, I.~Uschmann, E.~F\"orster, V.~Große, M.~Thuerk,
  T.~Koettig, F.~Schmidl, and P.~Seidel, ``The cubic to tetragonal phase
  transition in srtio$_3$ single crystals near its surface under internal and
  external strains,'' {\em Appl. Phys. Lett.}, vol.~96, p.~071901, 2010.

\bibitem{Cowley}
R.~A. Cowley, ``Lattice dynamics and phase transitions of strontium titanate,''
  {\em Phys. Rev.}, vol.~134, p.~A981, 1964.

\bibitem{Lytle}
F.~W. Lytle, ``X–ray diffractometry of low–temperature phase
  transformations in strontium titanate,'' {\em J. Appl. Phys.}, vol.~35,
  p.~2212, 1964.

\bibitem{Sanchez}
J.~C.~R. Sanchez, B.~Nelson-Cheeseman, M.~Granada, E.~Arenholz, and L.~B.
  Steren, ``Exchange-bias effect at la$_{0.75}$sr$_{0.25}$mno$_3$/lanio$_3$
  interfaces,'' {\em Phys. Rev. B}, vol.~85, p.~094427, 2012.

\bibitem{Peng}
J.~Peng, C.~Song, F.~Li, B.~Cui, H.~Mao, Y.~Wang, G.~Wang, and F.~Pan, ``Charge
  transfer and orbital reconstruction in strain-engineered
  (la,sr)mno$_3$/lanio$_3$ heterostructures,'' {\em ACS Appl. Mater.
  Interfaces}, vol.~7, p.~17700, 2015.

\bibitem{Celinski}
Z.~Celinski, K.~B. Urquhart, and B.~Heinrich, ``Using ferromagnetic resonance
  to measure the magnetic moments of ultrathin films,'' {\em J. Magn. Magn.
  Mater.}, vol.~166, p.~6, 1997.

\bibitem{Liu2}
M.~Liu, Z.~Zhou, T.~Nan, B.~M. Howe, G.~J. Brown, and N.~X. Sun, ``Voltage
  tuning of ferromagnetic resonance with bistable magnetization switching in
  energy‐efficient magnetoelectric composites,'' {\em Adv. Mater.}, vol.~25,
  p.~1435, 2013.

\bibitem{Liu3}
T.~Liu, H.~Chang, V.~Vlaminck, Y.~Sun, M.~Kabatek, A.~Hoffmann, L.~Deng, and
  M.~Wu, ``Ferromagnetic resonance of sputtered yttrium iron garnet nanometer
  films,'' {\em J. Appl. Phys.}, vol.~115, p.~17A501, 2014.

\bibitem{Brataas}
A.~Brataas, Y.~Tserkovnyak, and G.~E.~W. Bauer, ``Spin-pumping in
  ferromagnet–normal metal systems,'' {\em J. of Mag. and Mag. Mat.},
  vol.~272, p.~1981, 2004.

\bibitem{Mosendz3}
O.~Mosendz, G.~Woltersdorf, B.~Kardasz, B.~Heinrich, and C.~H. Back,
  ``Magnetization dynamics in the presence of pure spin currents in magnetic
  single and double layers in spin ballistic and diffusive regimes,'' {\em
  Phys. Rev. B}, vol.~79, p.~224412, 2009.

\bibitem{Du}
C.~Du, H.~Wang, F.~Yang, and P.~C. Hammel, ``Enhancement of pure spin currents
  in spin pumping y$_3$fe$_5$o$_{12}$/cu/metal trilayers through spin
  conductance matching,'' {\em Phys. Rev. Appl.}, vol.~1, p.~044004, 2014.

\bibitem{Deorani}
P.~Deorani and H.~Yang, ``Role of spin mixing conductance in spin pumping:
  Enhancement of spin pumping efficiency in ta/cu/py structures,'' {\em Appl.
  Phys. Lett.}, vol.~103, p.~232408, 2013.

\bibitem{Wei}
W.~Zhang, M.~B. Jungsfleisch, W.~Jiang, J.~Sklenar, F.~Y. Fradin, J.~E.
  Pearson, J.~B. Ketterson, and A.~Hoffmann, ``Spin pumping and inverse spin
  hall effects—insights for future spin-orbitronics (invited),'' {\em J.
  Appl. Phys.}, vol.~117, p.~172610, 2015.

\bibitem{Harii}
K.~Harii, Z.~Qiu, T.~Iwashita, Y.~Kajiwara, K.~Uchida, K.~Ando, T.~An,
  Y.~Fujikawa, and E.~Sautoh, ``Spin pumping in a
  ferromagnetic/nonmagnetic/spin-sink trilayer film: Spin current
  termination,'' {\em Key. Eng. Mater.}, vol.~508, p.~266, 2012.

\bibitem{Azevedo}
A.~Azevedo, L.~H. Vilela-Leao, L.~Rodr\'{i}guez-Su\'{a}rez, A.~F.~L. Santos,
  and S.~M. Rezende, ``Spin pumping and anisotropic magnetoresistance voltages
  in magnetic bilayers: Theory and experiment,'' {\em Phys. Rev. B}, vol.~83,
  p.~144402, 2011.

\bibitem{Mecking}
N.~Mecking, Y.~S. Gui, and C.-M. Hu, ``Microwave photovoltage and
  photoresistance effects in ferromagnetic microstrips,'' {\em Phys. Rev. B},
  vol.~76, p.~224430, 2007.

\bibitem{Obstbaum}
M.~Obstbaum, M.~H\"artinger, H.~G. Bauer, T.~Meier, F.~Swientek, C.~H. Back,
  and G.~Woltersdorf, ``Inverse spin hall effect in
  ni$_{81}$fe$_{19}$/normal-metal bilayer,'' {\em Phys. Rev. B}, vol.~89,
  p.~060407, 2014.

\bibitem{Ando2}
K.~Ando and E.~Saitoh, ``Observation of the inverse spin hall effect in
  silicon,'' {\em Nat. Commun.}, vol.~3, p.~629, 2012.

\end{thebibliography}

\end{document}